\documentclass[nobm]{epl}

\title{
Interface roughening in Hele--Shaw flows with quenched disorder:
experimental and theoretical results }
\shorttitle{
Interface roughening in Hele--Shaw flows}

\author{A. Hern\'{a}ndez-Machado\inst{1}
\thanks{E-mail: \email{aurora@ecm.ub.es}}
\and J. Soriano\inst{1}
\and A. M. Lacasta\inst{2} \and
M.A.  Rodr\'{\i}guez\inst{3} \and
L. Ram\'{\i}rez-Piscina\inst{2} \and
J. Ort\'{\i}n\inst{1}}
\shortauthor{A. Hern\'{a}ndez-Machado\etal}
\institute{
\inst{1}
Departament ECM, Facultat de F\'{\i}sica, Universitat de
Barcelona\\
Diagonal 647, E-08028 Barcelona, Spain \\
\inst{2}
Departament de F\'{\i}sica Aplicada,
Universitat Polit\`{e}cnica de Catalunya\\
Avda. Dr. Mara\~{n}on 44, E-08034 Barcelona, Spain.\\
\inst{3}
Instituto de F\'{\i}sica de Cantabria, CSIC\\
Avenida Los Castros, E-39005 Santander, Spain.}

\begin{document}

\maketitle

\begin{abstract}
We study the forced fluid invasion of an air-filled model porous medium
at constant flow rate, in 1+1 dimensions, both experimentally and
theoretically.  We focus on the non-local character of the interface
dynamics, due to liquid conservation, and its effect on the scaling
properties of the interface upon roughening.  Specifically, we study the
limit of large flow rates and weak capillary forces.  Our theory
predicts a roughening behaviour characterized at short times by a
growth exponent $\beta_1 = 5/6$, a
roughness exponent $\alpha_1=5/2$, and a dynamic exponent $z_1=3$, and
by $\beta_2=1/2$, $\alpha_2=1/2$, and $z_2=1$ at long times, before
saturation.  This theoretical prediction is in good agreement with the
experiments at long times.The ensemble of experiments, theory, and
simulations provides evidence for a  new universality class of
interface roughening in $1+1$ dimensions.
\end{abstract}

Studies on the morphology of interfaces moving in disordered media under
nonequilibrium conditions constitute an active field of research
\cite{meakin-nelson-kardar}.  One relevant example is the forced
invasion of an air-filled porous medium by a viscous wetting fluid such
as oil or water, which gives rise to a nonequilibrium rough interface
\cite{rubio,horvath,he,wong,delker}.  The roughening process of an
initially flat interface is described in terms of the interfacial
root-mean-square width $w$.  In many systems $w$ follows a Family-Vicsek
dynamical scaling \cite{family2}:  $w \sim t^{\beta}$ for
$t<t_{\times}$, $w \sim L^\alpha$ for $t>t_{\times}$ and $t_{\times}
\sim L^z$, with $\alpha = z \beta$.  Here $t_{\times}$ is a saturation
time, $\beta$ the growth exponent, $\alpha$ the roughness exponent, and
$z$ the dynamic exponent.  The roughness exponent can also be obtained
from the power spectrum $S(q,t)$, which is less sensitive to finite-size
effects.  We have \cite{miguel} $ S(q,t) = q^{-(2\alpha + 1)} s_{FV}(q
t^{1/z}) $, where $s_{FV}$ obeys $ s_{FV}(u) \sim constant $ when $u \gg
1$, and $ s_{FV}(u) \sim u^{2 \alpha +1} $ when $u \ll 1$.

Experiments of forced fluid invasion (FFI) in air-filled packings of
glass beads, in 1+1 dimensions, have given roughness exponents in the
range $\alpha = 0.6-0.9$ \cite{rubio,horvath,he,wong}.  The dispersion
reflects that the effective values of $\alpha$, determined by measuring
widths over several scales at saturation, depend on the capillary number
$C_a$ \cite{he}.  On the other hand, there are few experimental
determinations of the exponent $\beta$, because the large relative strength of
capillary to viscous forces made the growth regime extremely short in most
experiments.

In this letter we report new experimental results of dynamic interfacial
roughening in FFI, in 1+1 dimensions, at {\it constant injection rate}.
In our case the model porous medium is a Hele-Shaw cell with controlled
spatial fluctuations of the gap thickness.  While the viscous pressure
field and the surface tension in the plane of the cell keep the
interface smooth on large length scales, fluctuations in the gap
thickness produce local fluctuations in capillary pressure which roughen
the interface.  Thus the governing physics is the same than in a real
two-dimensional porous medium, and we have the possibility of
controlling the statistical properties and the relative strength of the
disorder.

\begin{figure}
\onefigure[width=8cm]{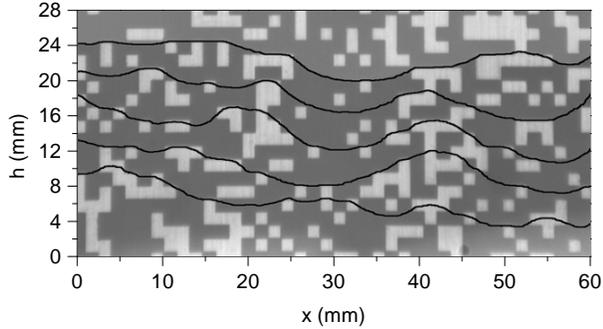}
\caption{Several consecutive close-up views
of a piece of the oil-air interface, taken at 150 s time
intervals. The flow rate is $Q=10$ ml/h.}
\label{Fig:fronts}
\end{figure}

In our experiments a silicone oil \cite{oil} displaces air in a
horizontal Hele-Shaw cell, $190 \times 550$ mm$^2$, made of two glass
plates 19 mm thick. Fluctuations in the gap thickness are
provided by a fiber
glass substrate which contains a random distribution of square copper
islands filling 35\% of the substrate.  Each island is 0.06 $\pm $
0.01 mm thick and has 1.500 $\pm$ 0.005 mm lateral size (Fig.\
\ref{Fig:fronts}).  The gap spacing from the substrate to the top plate
is $b = $ 0.36 $\pm $ 0.05 mm.
Since the maximum width of the meniscus is only one half the gap
width, the interface can be considered one--dimensional at the length
scale of the copper islands.
The oil is injected at one side of the cell using a constant flow-rate
syringe pump.  A flat front is first prepared by keeping
the oil at rest in a transverse copper track separated 2 mm from the
beginning of the noise.  Initially the oil is pushed at high flow rate
(300 ml/h) for about 4 s to keep the interface smooth while it
enters the noise. Next, at a time defined as $t=0$,  the flow rate is set
to its nominal value
$Q$, and the oil-air interface is monitored as a function
of time using two CCD cameras, with a final spatial resolution of 0.37
mm/pixel.
This imposes a cut--off at small length scales, about twice
the maximum width of the meniscus.
We choose a range of flow rates ($10 \le Q \le 100$ ml/h)
such that the contact line overcomes the copper islands rather easily,
without pinning or forming overhangs. For $Q = 10$ ml/h
(Fig.\ \ref{Fig:fronts}) the average interface velocity is $v=0.038
\pm 0.005$ mm/s, which we take as a reference.

\begin{figure}[tbh]
\onefigure[width=14cm]{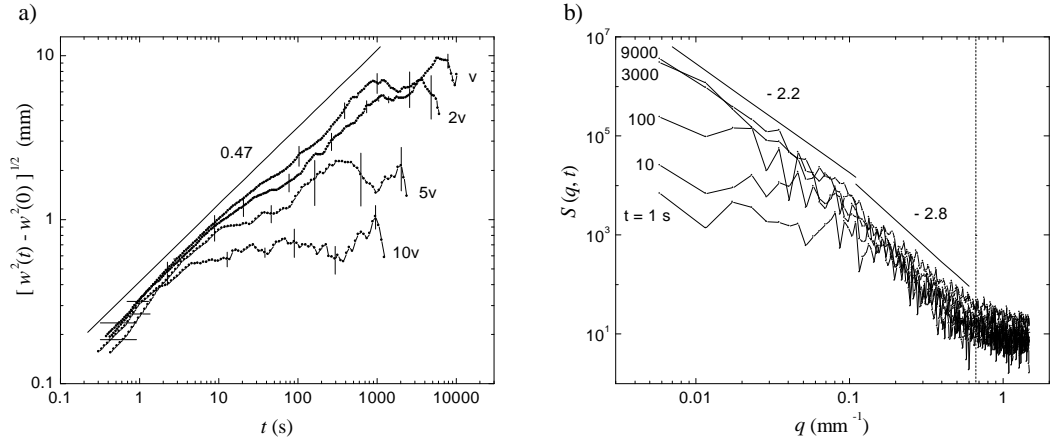}
\caption{
a) Double-logarithmic plot of $w(t)$ for four different
interface velocities. Each curve is an average over 6 experiments
(3 runs on 2 different realizations of the noise).
In each experiment $w^2(0)$ has been substracted from $w^2(t)$, in order
to minimize the effect of the initial condition. The error bars give a
measure of the overall dispersion between different experiments.
b) Power spectrum of the interface profiles at successive time
intervals, for the experiments at velocity $v$.
The vertical line indicates the spatial scale of the disorder.
}
\label{Fig:w-S-exp}
\end{figure}

Our results for $w(t)$ are presented in Fig.\ \ref{Fig:w-S-exp}(a).
As expected, both the saturation time and the interfacial width
$w$ at saturation depend on the average velocity at which the interface is driven.
We observe also large fluctuations in the interfacial width, consistent
with the observations reported in \cite{he}.
The fluctuations in $w$ are highly reproducible in experimental
runs with the same realization of the disorder.
Disregarding fluctuations, the growth of the interfacial
width is consistent with a power-law
$w \sim t^{\beta}$ with $\beta = 0.47 \pm 0.04$,
extending about three orders of magnitude in $t$ (for
measurements at velocity $v$) until the width saturates.

The power spectrum for the set of measurements at velocity $v$
is shown in Fig.\ \ref{Fig:w-S-exp}(b). At short times the
spectrum displays a {\it plateau} for small $q$ and a
power law decay at large $q$ as expected. As the interface advances
in the disorder the value of the {\it plateau} increases, and
we observe the emergence of another
scale-independent (power-law) behaviour for
an increasingly growing range of spatial scales, in agreement with the
scaling of $S(q,t)$ expected for $u=qt^{1/z} \ll 1$.
This behaviour is also observed at velocity $2v$. The measured
exponents are $-2.2 \pm 0.2$ ($\alpha = 0.6 \pm 0.1$) at small $q$
(long length scales) and $-2.8 \pm 0.2$ ($\alpha = 0.9 \pm 0.1$) at
large $q$ (short length scales).
The spectra of the experimental interfaces have been calculated imposing
periodic boundary conditions. To this purpose, we have substracted from the
interface the straight line joining its two ends. This correction
eliminates an artificial slope that is specially important at
saturation, where the difference in height between the two ends
is maximum \cite{periodic-boundary-conditions}.

One interesting aspect of roughening in fluid flows is the {\it
nonlocal} nature of the interfacial dynamics, due to liquid
conservation.  The importance of nonlocality in this problem has been
already pointed out by a number of authors \cite{he,delker,krug-meakin}.
Very recently this issue has been addressed explicitly in the
theoretical formulation of FFI at constant pressure \cite{ganesan} and of
spontaneous imbibition \cite{ganesan,nissila}.  In this letter we
address the same issue for FFI at constant flow rate.  This driving
condition has not been explored theoretically yet, in spite of being
often used in experiments.

Our model is based on a time dependent Ginzburg Landau model with
conserved order parameter (model B in Ref.\ \cite{hohenberg}):
\begin{eqnarray}
\frac{\partial \phi}{\partial t} = \nabla M \nabla
(-\phi +\phi^3- \epsilon^2 \nabla^2 \phi).
\label{modB}
\end{eqnarray}
$\phi$ is the order parameter with equilibrium values $\phi_{eq}=\pm 1$,
and $\epsilon$ is the interfacial width.
$M$ is the mobility, which we consider fluctuating in space:
\begin{eqnarray}
M(\phi,{\bf r})=
\left \{ \begin{array}{ll}
K ( 1+ \xi({\bf r})) & \mbox{for oil} ~~ (\phi>0) \\
0 & \mbox{for air} ~~ (\phi < 0)
\end{array} \right.
\label{mobility}
\end{eqnarray}
with $K$ a macroscopic mobility and
$\xi({\bf r})$ a weak static disorder with a spatial correlation:
\begin{eqnarray}
<\xi({\bf r})\xi({\bf r'})> =
2 D C \left(\frac{|{\bf r}-{\bf r'}|}{\lambda} \right ),
\label{correl}
\end{eqnarray}
where $\lambda$ is the disorder correlation lenght and $C$ is normalized.
Dub\'e {\it et al.} \cite{nissila} have proposed a similar formulation
for spontaneous imbibition.  In their problem the interface is driven
only by capillary forces, and moves with average velocity $\bar{v} \sim
t^{-1/2}$.  Here, instead, nonlocal viscous forces are dominant over the
local fluctuations of capillary pressure:  our interface is driven by a
constant flux $\gamma_0$, which gives rise to a steady state rough
interface moving with constant noise-dependent $\bar{v}$.  The disorder
is also introduced differently in the two approaches.  In our
two-dimensional model the fluctuations in gap thickness are represented
(phenomenologically) by a fluctuating mobility.

The macroscopic limit of the model is obtained by an asymptotic
expansion in orders of $\epsilon$, following a procedure described in
Ref.\ \cite{mozos}:
\begin{eqnarray}
&\nabla (1+\xi({\bf r})) \nabla P = 0 ~,\nonumber \\
&P = -\Gamma \kappa ~,
\label{macro2} \\
&v_n =  -  K ( 1 + \xi(x,y_{int})) (\nabla P)_n  ~,
\nonumber
\end{eqnarray}
where the pressure field in the bulk of the liquid is given by $P=
(\phi({\bf r},t) - \phi_{eq})/(2 \phi_{eq})$, the surface tension is
given by $\Gamma=(2 \phi_{eq})^{-2} \int dy (\partial_y \phi_{st})^2$,
$\phi_{st}$ is the steady planar solution of Eq.\ (\ref{modB}), $\kappa$
is the curvature of the interface, and $v_n$ its normal velocity.
$\xi(x,y_{int})$ is taken at the interface.
The deterministic part of Eq.\ (\ref{macro2})
reproduces the macroscopic equations for oil-air
displacements in a Hele-Shaw cell \cite{saffman}.

The interfacial equation in Fourier space can be derived from Eq.\
(\ref{macro2}) using Green function analysis.
We define ${\bar P} = P + \bar{\gamma}_0/K  ( y-\bar{\gamma}_0 t)$,
where $\bar{\gamma}_0=\gamma_0/(4 \phi_{eq})$.
In this case, the Green function $G({\bf r},{\bf r'})$
obeys $\nabla^2 G=\delta({\bf r}-{\bf r'}) $.
The interfacial equation
for the local deviations of the interface from its
mean position, $h(x,t)=y_{int} - \bar{v} t$, reads (to first order in
$\xi$):
\begin{eqnarray}
\int ds' G(s,s') \frac{v_n(s')-\bar{\gamma_0} \hat{n}\hat{y}}{K} =
\frac{1}{2} \left[- \Gamma \kappa(s)
+ \frac{\bar{\gamma}_0}{K} \left( h(s) + ({\bar v} - \bar{\gamma}_0) t
\right) \right]
\nonumber \\
+ \int ds' \hat{n} \nabla G(s,s') \left[ - \Gamma \kappa(s')
+ \frac{\bar{\gamma}_0}{K} \left( h(s')
+ ({\bar v} - \bar{\gamma}_0) t \right) \right]
+ \int ds' G(s,s')\frac{v_n}{K}
\xi(s',y_{int}),
\label{inter}
\end{eqnarray}
where $s$ is the contour variable on the interface and $\hat{n}$ and
$\hat{y}$ are the unitary vectors perpendicular to the interface and
along the $y$ direction respectively.  We have neglected volume terms
which do not contribute to the scaling.  A distinguishing feature of Eq.
(\ref{inter}) is the presence of integral terms, which
account for the {\it nonlocal} character of the interfacial dynamics.
The deterministic
part is equivalent to the equation for unstable displacements (air
displacing oil, $\bar{\gamma}_0 < 0$) derived in Refs.\
\cite{jasnow,langer}.

In this problem, due to the presence of a correlated disorder,
two different time scales of dynamic origin are present.
One of them is related to the ballistic dynamics of the interface due
to the driving and corresponds to the time in which the interface advances
the noise characteristic length in the $y$-direction,
$t_0 = \lambda/ \bar{\gamma}_0$.
The other temporal scale corresponds to the dynamics that couples
different points of the interface. For scales not much greater than $\lambda$
this dynamics is diffusive. The associated temporal scale
is the time required by the diffusion along the interface to reach distances
of the scale of the disorder, and is given by
$t_D = \lambda^2 / K$.
Here, we consider $t_0 \gg t_D$. This condition is fulfilled if
$\lambda \ll K / \bar{\gamma}_0$.
In this limit, the noise appears as persistent in the $y$-direction at scales
comparable to $\lambda$.
Therefore, we will take it as equivalent to a columnar (only $x$-dependent)
noise.
In this case, the stationary velocity ${\bar v}=\bar{\gamma}_0 + \left(
\bar{\gamma}_0/L \right) \int \xi(x) dx$, where $L$ is the system size
in the $x$ direction.

In Fourier space, for small deviations from a flat interface
($| q |h \ll 1$) we have:
\begin{eqnarray}
\frac{\partial \tilde{h}_q}{\partial t}
 = -  K  \Gamma q^2 |q| \tilde{h}_q
 - \bar{\gamma}_0 |q| \tilde{h}_q
 +  \bar{\gamma}_0 \xi_q (1- \delta_{q,0}).
\label{fourier}
\end{eqnarray}
We observe that short wavelength fluctuations are
damped by surface tension (first term of the r.h.s.),
while long wavelength fluctuations are damped by the advancement
of the front, driven by the external flux (second term of the r.h.s.).
Eq.\ (\ref{fourier}) leads to two different temporal
regimes.  The early growth regime is dominated by the first term of
(\ref{fourier}), and the dynamic exponent is
$z_1=3$. By linear scaling analysis we obtain $ z- \alpha - 1/2 =0$,
which gives $ \alpha_1=5/2$ and $ \beta_1=5/6$. Since $\alpha_1 > 1$, the
interface is superrough in the early time stages.
At longer times the second term of (\ref{fourier})
dominates the dynamics, and we get $z_2=1$,
$\alpha_2=1/2$, and $\beta_2=1/2$. The characteristic crossover time
between the two growth regimes is given by $t_c \sim
\zeta_c^3 \sim \left( K \Gamma / \bar{\gamma}_0 \right)^{3/2}$.

\begin{figure}[bth]
\onefigure[width=14cm]{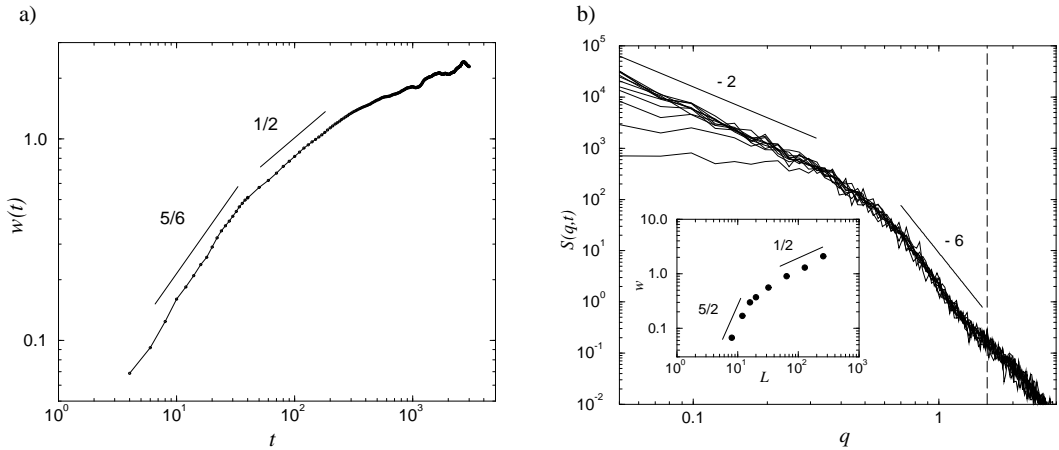}
\caption{
a) Temporal evolution of the interfacial width, from the numerical
integration of Eqs.\ (\protect{\ref{modB}})-(\protect{\ref{mobility}}),
for a system of size $L=256$.
b) Power spectrum at different time intervals,
for the same case of (a).
The vertical line is the spatial scale of the disorder. The
interfacial width at saturation vs.
the system size $L$ is shown in the inset.
}
\label{w-S-sim}
\end{figure}

We have checked these predictions by numerical integration of Eqs.\
(\ref{modB})-(\ref{mobility}) in a rectangular lattice with periodic
boundary conditions in the $x$-direction.  A constant flux $\gamma_0$
has been maintained by fixing the value of $\phi$ at a certain distance
behind the interface (typically 20 space units).
To mimic the experiments, we have assumed that
the oil mobility can take two different values, so that
$\xi$ is a dichotomous noise of values
$\pm D$.  The noise is defined in boxes of side $l_d=4$ space units,
randomly distributed with probability $0.35$ for the positive value.
The values of $l_d$ and $\gamma_0$ satisfy the requirement
of noise persistence in the $y$ direction.  Our results for
$\gamma_0=0.05$, $K=1$, $\epsilon=1$ and $D=0.5$, averaged over $25$
realizations of the noise, are presented in Figs.\ \ref{w-S-sim}(a) and
\ref{w-S-sim}(b).

Fig.\ \ref{w-S-sim}(a) is a log-log plot of $w(t)$, including the
analytical prediction for the growth exponents.  The numerical results
are consistent with a crossover between $\beta_1 = 5/6$ at short times
and $\beta_2 = 1/2$ at longer times.  Fig.\ \ref{w-S-sim}(b) shows a
log-log plot of $S(q,t)$.  At short times the
spectrum displays a {\it plateau} for small $q$, followed by a power law
corresponding to $\alpha_1 \simeq 5/2$ for large $q$.  At longer times,
$S(q,t)$ shows a data collapse at small $q$ into a power law
corresponding to $\alpha_2 \simeq 1/2$.  The same results are obtained
from the analysis of $w$ vs. system size $L$, shown in the inset of
Fig.\ \ref{w-S-sim}(b).  We observe the predicted crossover between
the two growth regimes.  Using the scaling relation $\alpha=z \beta$, the
dynamic exponents are $z_1 \simeq 3$ at short times and $z_2 \simeq 1$
at long times, in agreement with the analytical results.

Although we have experimental indications of very fast growth in the
earliest time stages, we have not been able to identify experimentally
the short--time scaling regime predicted by the model.  This regime is
difficult to access experimentally because it is very short, it is
obscured by the transient originated by setting $Q$ to its nominal
value, and, in addition, the short-time behaviour of $w(t)$ is strongly
dependent on the definition of $t=0$.  Nevertheless, the experimental
power spectrum shows that the short length scales saturate with a larger
roughness exponent than the long length scales, which points to a
different mechanism at the two scales, in the same direction as the
model.  The experimental and calculated values of the roughness
exponent at short length scales do not coincide because the details of
the physical mechanisms operative at the shortest length scales
(capillary phenomena) are not properly captured by the model.

Concerning the long time and long length
scaling regime, several authors have argued
that FFI should fall in the KPZ universality class ($\beta = 1/3$ and
$\alpha = 1/2$ in 1+1 dimensions) in the limit of large $C_a$ considered
here \cite{kessler,he}.  Our experimental and theoretical results
contradict this prediction for the exponent $\beta$, which is
consistently found to be $\beta \simeq 1/2$.
Concerning the roughening exponent $\alpha$, however, previous experimental
measurements \cite{he} and our own experiments give $\alpha > 0.6$
or larger.  Although the value of $\alpha$ decreases with increasing
$C_a$, the theoretical limit $\alpha = 0.5$ is presumably not accessible
because the saturation width $w$ falls to magnitudes comparable to the
gap thickness or to the typical pore size as $C_a$ becomes very large.

In conclusion, we have studied the dynamics of interfacial roughening in
forced fluid invasion of a model porous medium at constant flow rate, in
1+1 dimensions.  We have focused on displacements in which the viscous
pressure field dominates over the fluctuations in capillary pressure
(large $C_a$, weak disorder), and found a new universality class with
two distinct time regimes.

\acknowledgments

We acknowledge the financial support of the Direcci\'on General
de Ense\~nanza Superior (Spain) under projects
BFM2000-0628-C03-01 and BFM2000-0624-C03-02.


\begin{thebibliography}{0}
\bibitem{oil}
The silicone oil has kinematic viscosity
$\nu =50$ mm$^2$/s, density $\rho =0.99$ g/cm$^3$, and surface tension
with air $\sigma =20.7$ mN/m at room temperature.
\bibitem{periodic-boundary-conditions}
Although the different height at the end points should ideally be
attributed to long wavelength fluctuations (small $q$), we believe that in our
experimental setup it is dominated by the wetting of the gap
spacers by the invading oil, which changes the physics at the two ends.
For this reason, we have ignored 8 mm of the interface at each of the
two ends, and imposed the linear correction explained in the text.
Measurements of $w$ as a function of the size $l$ of a local window
demonstrate that the correction does not modify the measured exponents.
\end{thebibliography}
\end{document}